\documentstyle[11pt,epsfig]{article}
\newcounter{subequation}[equation]

\makeatletter

\def\thesubequation{\theequation\@alph\c@subequation}
\def\@subeqnnum{{\rm (\thesubequation)}}
\def\slabel#1{\@bsphack\if@filesw {\let\thepage\relax
   \xdef\@gtempa{\write\@auxout{\string
      \newlabel{#1}{{\thesubequation}{\thepage}}}}}\@gtempa
   \if@nobreak \ifvmode\nobreak\fi\fi\fi\@esphack}
\def\subeqnarray{\stepcounter{equation}
\let\@currentlabel=\theequation\global\c@subequation\@ne
\global\@eqnswtrue
\global\@eqcnt\z@\tabskip\@centering\let\\=\@subeqncr
$$\halign to \displaywidth\bgroup\@eqnsel\hskip\@centering
  $\displaystyle\tabskip\z@{##}$&\global\@eqcnt\@ne
  \hskip 2\arraycolsep \hfil${##}$\hfil
  &\global\@eqcnt\tw@ \hskip 2\arraycolsep
  $\displaystyle\tabskip\z@{##}$\hfil
   \tabskip\@centering&\llap{##}\tabskip\z@\cr}
\def\endsubeqnarray{\@@subeqncr\egroup
                     $$\global\@ignoretrue}
\def\@subeqncr{{\ifnum0=`}\fi\@ifstar{\global\@eqpen\@M
    \@ysubeqncr}{\global\@eqpen\interdisplaylinepenalty \@ysubeqncr}}
\def\@ysubeqncr{\@ifnextchar [{\@xsubeqncr}{\@xsubeqncr[\z@]}}
\def\@xsubeqncr[#1]{\ifnum0=`{\fi}\@@subeqncr
   \noalign{\penalty\@eqpen\vskip\jot\vskip #1\relax}}
\def\@@subeqncr{\let\@tempa\relax
    \ifcase\@eqcnt \def\@tempa{& & &}\or \def\@tempa{& &}
      \else \def\@tempa{&}\fi
     \@tempa \if@eqnsw\@subeqnnum\refstepcounter{subequation}\fi
     \global\@eqnswtrue\global\@eqcnt\z@\cr}
\let\@ssubeqncr=\@subeqncr
\@namedef{subeqnarray*}{\def\@subeqncr{\nonumber\@ssubeqncr}\subeqnarray}
\@namedef{endsubeqnarray*}{\global\advance\c@equation\m@ne%
                           \nonumber\endsubeqnarray}

\makeatletter \@addtoreset{equation}{section} \makeatother
\renewcommand{\theequation}{\thesection.\arabic{equation}}
\font\elevenmsb=msbm10 scaled 1100

\def\dalemb#1#2{{\vbox{\hrule height .#2pt
        \hbox{\vrule width.#2pt height#1pt \kern#1pt
                \vrule width.#2pt}
        \hrule height.#2pt}}}
\def\square{\mathord{\dalemb{6.8}{7}\hbox{\hskip1pt}}}

    \let\e=\epsilon
  \let\q=\theta  
  \let\n=\nu

 \def\bd{\begin{document}} \def\ed{\end{document}}
\def\ds{\documentstyle} \let\fr=\frac \let\bl=\bigl \let\br=\bigr
\let\Br=\Bigr \let\Bl=\Bigl
\let\bm=\bibitem
\let\na=\nabla
\let\pa=\partial \let\ov=\overline
\def\ie{{\it i.e.\ }}
\newcommand{\be}{\begin{equation}}
\newcommand{\ee}{\end{equation}}
\def\ba{\begin{array}}
\def\ea{\end{array}}
\def\ft#1#2{{\textstyle{{\scriptstyle #1}\over {\scriptstyle #2}}}}
\def\fft#1#2{{#1 \over #2}}
\def\del{\partial}
\def\sst#1{{\scriptscriptstyle #1}}
\def\oneone{\rlap 1\mkern4mu{\rm l}}
\def\e7{E_{7(+7)}}
\def\td{\tilde}
\def\wtd{\widetilde}
\def\im{{\rm i}}
\def\bog{Bogomol'nyi\ }
\def\q{{\tilde q}}
\def\hast{{\hat\ast}}
\def\0{{\sst{(0)}}}
\def\1{{\sst{(1)}}}
\def\2{{\sst{(2)}}}
\def\3{{\sst{(3)}}}
\def\4{{\sst{(4)}}}
\def\5{{\sst{(5)}}}
\def\6{{\sst{(6)}}}
\def\7{{\sst{(7)}}}
\def\8{{\sst{(8)}}}
\def\n{{\sst{(n)}}}
\def\oo{{\"o}}
\def\hA{\hat{\cal A}}
\def\ns{{\sst {\rm NS}}}
\def\rr{{\sst {\rm RR}}}
\def\tH{{\widetilde H}}
\def\tB{{\widetilde B}}
\def\cA{{\cal A}}
\def\cF{{\cal F}}
\def\tF{{\wtd F}}
\def\Z{\rlap{\sf Z}\mkern3mu{\sf Z}}
\def\ep{{\epsilon}}
\def\IIA{{\rm IIA}}
\def\IIB{{\rm IIB}}
\def\ads{{\rm AdS}}
\def\R{\rlap{\rm I}\mkern3mu{\rm R}}
\def\mapright#1{\smash{\mathop{-\!\!\!-\!\!\!-\!\!\!-\!\!\!-\!\!\!
             \longrightarrow}\limits^{#1}}}

\def\Ei{{\hbox{Ei}}}
\def\Ci{{\hbox{Ci}}}
\def\Si{{\hbox{Si}}}

\newcommand{\ho}[1]{$\, ^{#1}$}
\newcommand{\hoch}[1]{$\, ^{#1}$}
\newcommand{\bea}{\begin{eqnarray}}
\newcommand{\eea}{\end{eqnarray}}
\newcommand{\ra}{\rightarrow}
\newcommand{\lra}{\longrightarrow}
\newcommand{\Lra}{\Leftrightarrow}
\newcommand{\aap}{\alpha^\prime}
\newcommand{\bp}{\tilde \beta^\prime}
\newcommand{\tr}{{\rm tr} }
\newcommand{\Tr}{{\rm Tr} }
\newcommand{\NP}{Nucl. Phys. }

\newcommand{\brussels}{\it Physique Th\'eorique et Math\'ematique,
Universit\'e Libre de Bruxelles,\\ Campus Plaine C.P. 231, B-1050
Bruxelles, Belgium}

\newcommand{\auth}{X. Bekaert\hoch{\clubsuit}, N. Boulanger\hoch{\sharp}
and J.F. V\'{a}zquez-Poritz\hoch{\natural}}

\begin{document}
\begin{flushright}
ULB-TH/02-17\\
June  2002\\
\hfill{\bf hep-th/0206050}\\
\end{flushright}

\begin{center}

{\large {\bf Gravitational Lorentz Violations from M-Theory}}

\vspace{20pt}

\auth

\vspace{10pt}
\brussels\\
\vspace{10pt}

\vspace{30pt}

\underline{ABSTRACT}
\end{center}

In an attempt to bridge the gap between M-theory and braneworld
phenomenology, we present various gravitational Lorentz-violating
braneworlds which arise from $p$-brane systems. Lorentz invariance
is still preserved locally on the braneworld. For certain
$p$-brane intersections, the massless graviton is {\sl
quasi}-localized. This also results from an M5-brane in a $C$-field. In
the case of a $p$-brane perturbed from extremality, the quasi-localized
graviton is massive. For a braneworld arising from {\sl global} $AdS_5$,
gravitons travel faster when further in the bulk, thereby apparently
traversing distances faster than light. 
{\vfill\leftline{} \vfill \vskip 10 pt \footnoterule {\footnotesize
\hoch{\clubsuit} \hoch{\sharp} \hoch{\natural} Supported in part by the
``Actions de Recherche Concert{\'e}es" of the ``Direction de la Recherche
Scientifique - Communaut\'e Francaise de Belgique," IISN-Belgium
(convention 4.4505.86).}\\ 
{\footnotesize \hoch{\clubsuit} \hoch{\sharp} Supported in part by a
``P\^ole d'Attraction Interuniversitaire" (Belgium) and the European
Commission RTN programme HPRN-CT-2000-00131, in which we are associated
with K.U. Leuven.}\\
{\footnotesize \hoch{\sharp} Supported in part by the ``Chercheur
F.R.I.A." (Belgium).}\\
{\footnotesize \hoch{\natural} Supported in part by the ``Francqui
Foundation" (Belgium).}

\vskip  -12pt}

\pagebreak \setcounter{page}{1}


\section{Introduction}

For certain non-factorizable five-dimensional geometries,
four-dimensional gravity is recovered at low-energy scales
\cite{randall2,randall1}. In particular, for a five-dimensional
spacetime with a metric given by
\be ds_5^2=a^2(z)(-dt^2+dx_i^2+dz^2), \label{5metric} \ee
then for certain factors $a(z)$, gravity can be localized on a
four-dimensional Minkowski surface of a domain wall with the
directions $t$, $x_i$. For example, the second Randall-Sundrum
braneworld scenario (RS2) used a warped five-dimensional metric that could
be identified as a slice of $AdS_5$, for which
$a(z)=(1+k|z|)^{-1}$. The absolute value ensures $\mbox{\elevenmsb
Z}_2$ symmetry about the domain wall, which corresponds to a
delta-function potential well which localizes the massless
graviton state\footnote{Since an M-theoretic origin for this
source is not known, a braneworld scenario which does not require
this is proposed in \cite{resolve}.}

Additional factors in the metric (\ref{5metric}) can yield two
types of deviations from four-dimensional Minkowski space embedded
in five dimensions. Firstly, additional factors could depend only
on our four-dimensional spacetime, such as in the case of a
cosmological spacetime on the braneworld. On the other hand, if
additional factors depend only on the bulk dimension, then the
spacetime on the braneworld will appear Minkowski for its
inhabitants but Lorentz invariance will be broken globally
\cite{Kalbermann,Chung,Ishihara,dan,csaki1,csaki2,csaki3}. The second
case will be the focus of this paper. The corresponding metric can
be written as\footnote{We do not include factors which would
globally break rotational invariance, as this could be in
contradiction with the isotropy of the cosmic microwave background
\cite{csaki1}.}
\be ds_5^2=a^2(z)(-b^2(z)\, dt^2+dx_i^2+dz^2). \label{asymm} \ee
Thus, the five-dimensional space is foliated by surfaces which
exhibit Poincar\'{e} invariance for fixed $z$, leading to
Lorentz-invariant Standard Model particle physics on the
four-dimensional braneworld, located at $z=0$. Gravitons, on the
other hand, are not constrained to the braneworld. Therefore,
gravitational processes may reflect the fact that four-dimensional
Lorentz symmetry is broken globally.

In particular, from (\ref{asymm}) the local speed of gravitational
propagation is given by $b(z).$\footnote{One can rescale $t$ in
order for the speed of light on the braneworld at $z=0$ to be
unity.} If $b(z)$ increases with $z$ then gravitational effects
may propagate from one region of our universe to another by
bending in the extra dimension, thereby traveling faster than
light on the braneworld \footnote{One recent proposal for
measuring the speed of propagation of gravity (over relatively short
distances) was given in \cite{experiment}.}. Thus, regions of the universe
not within each others' lightcones may actually affect each other via
gravitation, leading to an apparent violation of causality. This
suggested a non-inflationary solution to the cosmological horizon
problem \cite{Kalbermann,Chung,Ishihara}. It is important to keep
in mind that causality is still maintained from the
five-dimensional vantage point.

Using Birkhoff's theorem, the metric (\ref{asymm}) can always be
transformed into that of a black hole, assuming the simplest
possible sources in the bulk \cite{csaki1}. In particular, in the
case of a Reissner-Nordstrom black hole in $AdS_5$, the graviton
speed $b(z)$ increases away from the braneworld for a significant
fraction of the charge-mass parameter space \cite{csaki1, csaki2,
csaki3}. In a certain coordinate frame, the surface of the
braneworld is bent in the bulk due to the presence of the black
hole, while the graviton geodesics are straight lines. This is
another way of visualizing that the gravitons would go into the
bulk in order to connect two points which are on the braneworld.

We now ask the question: what M-theoretic factors give rise to a
braneworld which is bent in the bulk? Although M-theory motivates
the consideration of extra dimensions, it could be that braneworld
scenarios are theoretically complete within a five-dimensional
framework. However, certain aspects of braneworlds scenarios may
have natural M-theoretic origins. Bridging the gap between
M-theory and braneworld phenomenology may bring to light some
connections with M-theoretic ideas, which may not be apparent from
a five-dimensional vantage point. In this vein, we present various
gravitational Lorentz-violating braneworld scenarios which arise
naturally from $p$-brane systems. Although such scenarios
generally arise from five-dimensional black holes, a plethora of
higher-dimensional origins emerge within the framework of
M-theory.

This paper is organized as follows. In Section 2, we consider
braneworld scenarios which lift to the near-horizon region of
$p$-brane intersections. For the cases discussed, the spatial
portion of the braneworld corresponds to relative transverse
dimensions in the $p$-brane system, which results in Lorentz
invariance being globally broken. In Section 3, we find that a braneworld
scenario resulting from the near-horizon region of an M5-brane in a
constant $C$-field also exhibits the global breaking of Lorentz
invariance. For both scenarios, the massless graviton is $quasi$-localized
to the brane. In Section 4, we present an example of a globally
Lorentz-violating braneworld scenario arising from the near-horizon region
of a $p$-brane which is perturbed from extremality, and show that the
quasi-localized graviton is massive. In Section 5, we consider a
braneworld scenario which arises from $AdS_5$, which could be interpreted
as the near-horizon region of a D3-brane reduced on $S^5$. For {\sl
global} $AdS$, the breaking of global Lorentz invariance leads to
gravitons on the braneworld which, from our vantage point, traverse
distances faster than light. In section 6, we discuss some open issues.

\section{Braneworld from intersecting $p$-branes}

Albeit the imposed $\mbox{\elevenmsb Z}_2$ symmetry about $z=0$,
five-dimensional gravity-trapping domain walls arise from ten
dimensions as sphere reductions of the near-horizon region of
D$p$-branes for $p=3,4,5$. The domain wall resulting from a
D3-brane was originally used in the RS2 braneworld scenario while
the D4 and D5 branes yield dilatonic domain walls. In addition,
the D4 and M5 branes dimensionally reduce to the same domain wall,
as does the D5-brane and M5/M5-brane intersection \cite{cvetic}.

In the aforementioned cases, the observed four-dimensional
spacetime lies entirely within the (overall) worldvolume of the
$p$-brane(s). This is partially motivated by holography, since the
above cases for which gravity is localized all have a natural
decoupling limit, which suggests that the localization of gravity
may generally be realized within a Domain-wall/QFT
correspondence\footnote{The delta-function domain-wall source due
to the imposed $\mbox{\elevenmsb Z}_2$ symmetry provides an
ultra-violet cut-off in a dual quantum field theory.}. For
example, in the original RS2 scenario, the directions of the
effective four-dimensional universe correspond to the worldvolume
of a D3-brane in ten dimensions. If the worldvolume of the
D3-brane is Minkowski, then so is that of the braneworld, thus
yielding an effective four-dimensional universe that is strictly
Lorentz invariant. However, one could envision a system of
intersecting $p$-branes for which some of the resulting braneworld
directions do not lie within the overall worldvolume of the
$p$-branes. If all of the three observed spatial dimensions are
relative transverse to a $p$-brane, the result may be a braneworld
scenario which globally breaks gravitational Lorentz invariance,
while spatial rotational invariance is maintained.

\subsection{Braneworld from NS1/D2 system}

One way to achieve this is to add a pp-wave along the worldvolume
of a D3-brane in type IIB theory. The near-horizon limit of the
solution then becomes $K_5 \times S^5$, where $K_5$ is the
generalized Kaigorodov metric in $D=5$, and the geometry is dual
to a conformal field theory in the infinite momentum frame
\cite{boost}. T-dualizing along the direction of the pp-wave
results in the NS1/D2 system of type IIA theory. The corresponding
metric is given by
\be ds_{10}^2=H_1^{-3/4}H_2^{-5/8}\Big( -dt^2+ H_2
dx_1^2+H_1(dx_2^2+dx_3^2)+ H_1 H_2(dr^2+r^2 d\Omega_5^2)\Big)
,\label{metric} \ee
where
\be H_1=1+\frac{R_1^4}{r^4},\ \ \ \ \ \ \ H_2=1+\frac{R_2^4}{r^4},
\ee
and $R_1$ and $R_2$ are the charges of the NS1 and D2 brane,
respectively. This brane system can be illustrated by the
following diagram:

\bigskip\bigskip
\centerline{
\begin{tabular}{c|ccccccccccc}
&$t$ & $x_1$ & $x_2$ & $x_3$ & $r$ & $s_1$ & $s_2$ & $s_3$ & $s_4$
& $s_5$ & \\ \hline D2&$\times$ & $-$ & $\times$ & $\times$ & $-$
& $-$ & $-$ &
$-$ & $-$ & $-$ & $H_2$ \\
NS1&$\times$ & $\times$ & $-$ & $-$ & $-$ & $-$ & $-$ & $-$ &
$-$ & $-$ & $H_1$ \\
\end{tabular}}
\bigskip

\centerline{Diagram 1. The NS1/D2 system.}
\bigskip\bigskip

In the near-horizon region, we neglect the ``1" in the harmonic
functions and, for simplicity, consider the case $R_1=R_2\equiv
R$. The metric (\ref{metric}) can then be expressed as
\be ds_{10}^2=(1+k|z|)^{-3/2}\Big(
-(1+k|z|)^{-4}dt^2+dx_i^2+dz^2\Big) +(1+k|z|)^{1/2}R^2
d\Omega_5^2,\label{metric2} \ee
where
\be \frac{r}{R}=(1+k|z|)^{-1},\label{z} \ee
and $i=1,2,3$. We use the following metric Ansatz for dimensional
reduction over $S^n$
\be ds_D^2={\rm e}^{-2\alpha \phi}ds_d^2+g^{-2}{\rm
e}^{\frac{2(d-2)}{n}\alpha \phi}d\Omega_n^2,\label{ansatz} \ee
where $\alpha=-\sqrt{\frac{n}{2(D-2)(d-2)}}$. Reducing over $S^5$
in (\ref{metric2}) using the Ansatz (\ref{ansatz}) we obtain
\be ds_5^2=(1+k|z|)^{-2/3}\Big(
-(1+k|z|)^{-4}dt^2+dx_i^2+dz^2\Big). \label{metricd3} \ee
The fluctuations of the five-dimensional graviton satisfy the
equation of a minimally-coupled scalar field in a gravitational
background given by
\be
\partial_{\mu}\sqrt{-g}g^{\mu \nu}\partial_{\nu}\Phi=0.\label{min}
\ee
Taking $\Phi=M(t,x_i)\, (1+k|z|)^{3/2}\psi(z)$ we obtain a
Schr\"{o}dinger-type wave equation,
\be -\partial_z^2 \psi+V(z) \psi=m^2 \psi,\label{D3} \ee
with an effective potential given by
\be V(z)=\frac{15k^2}{4(1+k|z|)^2}-3\delta (z). \label{V} \ee
Note that the mass $m$ is defined by
\be \square_{(4)} M(t,x_i)=m^2 M(t,x_i), \ee
where $\square_{(4)}$ is the four-dimensional Laplacian.

The graviton wave equation (\ref{D3}) is the same as for a
braneworld originating from the D3-brane near-horizon region, such
as the original RS2 scenario. The massless graviton is localized
on the braneworld, as can by seen by the fact that the
corresponding wavefunction,
\be \psi=N(1+k|z|)^{-3/2}, \ee
is square-normalizable. The effective four-dimensional Newtonian
potential is modified in the same way, via the massive graviton
modes, as for braneworld arising from a D3-brane. In the present
case, however, Lorentz invariance is globally broken, as can be
seen from the form of the metric (\ref{metricd3}). The speed of
gravitons in the bulk is $v(z)=(1+k|z|)^{-2}$, where the speed of
light on the braneworld is unity. Since $v(z)$ decreases away from
the braneworld, the propagation of gravitons will not exhibit
effects that appear to violate causality. However, a
Lorentz-violating effect, from the point of view of the
braneworld, is a modification of the dispersion relation,
$E^2=m^2+c^2\, \vec{p}^2$, for which $c$ now depends on the spread
of the wave function in the extra dimension \cite{csaki1, rubakov,
coleman, burgess}.

The metric (\ref{metricd3}) is of the form (\ref{asymm}) where
$b(z) \longrightarrow 0$ as $|z| \longrightarrow \infty$. In
addition, $V(z)$ given by (\ref{V}) tends to a constant for large
$|z|$. This implies that the massless graviton is now {\sl
quasi}-localized for $\vec{p}\, ^2 >0$, that is, metastable
against escape into the extra dimension. In particular, the escape
width $\Gamma(|\vec{p} |)$ increases with $|\vec{p} |$
\cite{dubovsky}. This is true also for massless particles other
than the graviton, provided that the corresponding fields have
bulk modes. This type of metastability may have phenomenological
consequences, especially for ultra-high energy cosmic rays
\cite{rubakov, bertolami1, bertolami2}.

\subsection{Other examples}

We find similar results if we add a pp-wave to the worldvolume of
a D4 or D5-brane and then T-dualize the solution. In the case of a
D4-brane with a pp-wave, T-dualizing along the direction of the
wave yields the NS1/D3 system of type IIB theory. Note that this
is dual to the type IIA D2/D2 system, by lifting up to eleven
dimensions and the reducing along a different direction. The D2/D2
system can be represented by the following diagram:

\bigskip\bigskip
\centerline{
\begin{tabular}{c|ccccccccccc}
&$t$ & $x_1$ & $x_2$ & $x_3$ & $x_4$ & $r$ & $s_1$ & $s_2$ & $s_3$
& $s_4$ & \\ \hline D2&$\times$ & $-$ & $-$ & $\times$ & $\times$
& $-$ & $-$ &
$-$ & $-$ & $-$ & $H$ \\
D2&$\times$ & $\times$ & $\times$ & $-$ & $-$ & $-$ & $-$ & $-$ &
$-$ & $-$ & $K$ \\
\end{tabular}}
\bigskip

\centerline{Diagram 2. The D2/D2 system.}
\bigskip\bigskip

Without going into the details, in the near-horizon region we
reduce over the transverse $S^4$ and one of the $x_i$ directions
by using the metric Ansatz (\ref{ansatz}). After a coordinate
transformation, we obtain
\be ds_5^2=(1+k|z|)^{-4/3}\Big(
-(1+k|z|)^{-6}dt^2+dx_i^2+dz^2\Big) . \ee
Taking $\Phi=M(t,x_i)(1+k|z|)^{3/2}\psi(z)$ yields a
Schr\"{o}dinger-type wave equation,
\be
\partial_z^2 \psi+\Big( \frac{35k^2}{4(1+k|z|)^2}-5\delta (z)\Big)\psi=m^2
\psi. \label{D4} \ee
This is the identical graviton wave equation as for the braneworld
resulting from the D4-brane.

In a similar manner, T-dualizing a D5-brane with a pp-wave yields
the NS1/D4, which is dual to the D3/D3 system. This brane system
can be illustrated by the following diagram:

\bigskip\bigskip
\centerline{
\begin{tabular}{c|ccccccccccc}
&$t$ & $x_1$ & $x_2$ & $x_3$ & $x_4$ & $x_5$ & $r$ & $s_1$ & $s_2$
& $s_3$ & \\ \hline D3&$\times$ & $\times$ & $-$ & $-$ & $\times$
& $\times$ & $-$ &
$-$ & $-$ & $-$ & $H$ \\
D3&$\times$ & $\times$ & $\times$ & $\times$ & $-$ & $-$ & $-$ &
$-$ &
$-$ & $-$ & $K$ \\
\end{tabular}}
\bigskip

\centerline{Diagram 3. The D3/D3 system.}
\bigskip\bigskip

In the near-horizon region, a reduction over the transverse $S^3$
and two of the $x_i$ directions yields
\be ds_5^2={\rm e}^{-\frac{k}{3}|z|}(-{\rm e
}^{-k|z|}dt^2+dx_i^2+dz^2), \ee
after a coordinate transformation. Taking $\Phi=M(t,x_i)
(1+k|z|)^{3/2}\psi(z)$, we obtain a Schr\"{o}dinger-type wave
equation
\be
\partial_z^2 \psi+\Big( \frac{k^2}{4}-\delta (z)\Big)\psi=m^2
\psi. \label{D5} \ee
This is the same graviton wave equation as for the braneworld from
the D5-brane.

In all of these scenarios, the graviton speed decreases away from
the braneworld, so we would not observe causality violations
induced by bulk gravitational propagation. However, as previously
mentioned, the dispersion relation is modified and the massless
graviton is quasi-localized.

\section{Braneworld from M5-brane with $C$-field}

Consider the M5-brane solution with a constant $C$-field \cite{russo}: 
$$
ds_{11}^2=K^{1/3}H^{1/3}\Big(
\frac{1}{H}(-dt^2+dx_4^2+dx_5^2) +\frac{1}{K}dx_i^2+dr^2+r^2 d\Omega_4^2
\Big),
$$
\be 
H=1+\frac{R^3}{r^3},\quad K=\sin^2 \theta +\cos^2 \theta\, H, 
\ee
$$
dC_{(3)}=\sin \theta\, dH^{-1}\wedge dt\wedge dx_4\wedge dx_5+3R^3
\Omega_{(4)} \cos \theta -6\tan \theta\, dK^{-1}\wedge dx_1\wedge
dx_2\wedge dx_3, 
$$
where $i=1,2,3$. This can be interpreted as a 2-brane lying within a
5-brane \cite{russo2}. In the near-horizon region we can neglect the
``1" in $H$\, \footnote{In this decoupling limit, a holographic dual
theory would be non-associative.}. Reducing over $S^4$,
$x_4$ and $x_5$ yields a five-dimensional domain wall with the metric
\be 
ds_5^2=(1+k|z|)^{-10/3}\Big( [\cos^2 \theta +\sin^2 \theta\, (1+k|z|)^{-6}
](-dt^2+dz^2)+dx_i^2\Big),
\ee
where $r/R=(1+k|z|)^{-2}$. Taking $\Phi=M(t,x_i)(1+k|z|)^{5/2}\psi(z)$, we 
find that the graviton equation of motion can be expressed as 
\be 
-\partial_z^2 \psi+\Big( \frac{35k^2}{4(1+k|z|)^2}-5k\, \delta(z) \Big)
\psi=m^2 \Big( \cos^2 \theta +\frac{sin^2
\theta}{(1+k|z|)^6}\Big) \psi. 
\ee
The massless case coincides with the braneworld scenario arising from the
near-horizon of a D4-brane or M5-brane without a $C$-field, given by
(\ref{D4}) and studied in \cite{cvetic}. The graviton speed decreases away
from the braneworld, so we would not observe causality violations induced
by bulk gravitational propagation. However, as with the intersecting
$p$-brane origins discussed in the previous section, the dispersion
relation is modified and the massless graviton is quasi-localized.

\section{Braneworld from non-extremal $p$-brane}

The gravitational breaking of Lorentz invariance may correspond to
a perturbation away from extremality in the higher-dimensional
$p$-brane origin of the braneworld scenario. A case that is
tractable with semi-analytical methods \cite{jvp2} is the 5-brane
solution of heterotic or type II theories, whose metric is given
by
\be ds_{10}^2= H^{-1/4}(-f\, dt^2+dx_j^2)+H^{3/4}(f^{-1}dr^2+r^2
d\Omega_3^2), \label{D5metric} \ee
where
\be H=1+\frac{R^2}{r^2},\quad f=1-\frac{e^2 R^2}{r^2}, \ee
and $j=1,..,5.$ The non-extremality parameter is $e$. We suppose
that the brane is perturbed slightly from extremality so that $e
\ll 1$. In the near-horizon region we can neglect the ``1" in $H$,
and the metric (\ref{D5metric}) can be expressed as
\be ds_{10}^2=({\rm e}^{-k|z|}+e^2)^{1/4} \Big(
f(-dt^2+dz^2)+dx_j^2+R^2 d\Omega_3^2 \Big), \ee
where $r/R=\sqrt{{\rm e}^{-k|z|}+e^2}$. Reducing this region of
the 5-brane metric on $T^2 \times S^3$ with the Ansatz
(\ref{ansatz}) yields a five-dimensional domain wall with the
metric
\be ds_5^2=({\rm e}^{-k|z|}+e^2)^{2/3} \Big( (1+e^2\, {\rm
e}^{k|z|})^{-1} (-dt^2+dz^2)+dx_i^2 \Big), \ee
where $i=1,2,3$. The speed of gravitons in the bulk is
$v(z)=(1+e^2\, {\rm e}^{k|z|})^{-1}$. Since $v(z)$ decreases away
from the braneworld, this scenario will not exhibit
characteristics that would appear to violate causality from the
four-dimensional perspective. However, as we shall see, one
consequence of the global breaking of Lorentz invariance is that
the effective four-dimensional graviton is now a quasi-localized
massive state.
\begin{figure}
   \epsfxsize=4.0in
   \centerline{\epsffile{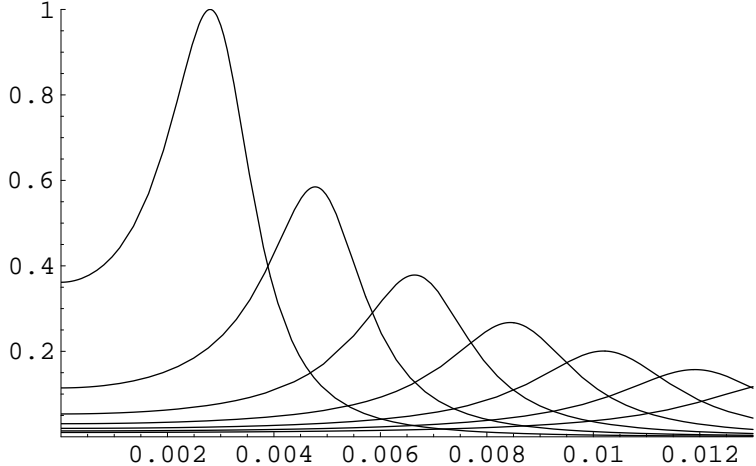}}
   \caption[FIG. \arabic{figure}.]{$|\psi (z=0)|^2$ versus $m/k$ for
$e=.005$, $.01$, $.015$, $.02$, $.025$, $.03$, and $.035$}
\end{figure}
Considering a graviton fluctuation which obeys the equation of
motion for a minimally-coupled scalar given by (\ref{min}), we
take $\Phi=({\rm e}^{-k|z|}+e^2)^{-1/2}M(t,x_i)\psi(z)$. The
resulting wave equation is
\be -\partial_z^2 \psi+\Big( \frac{\frac{k^2}{2}-m^2}{1+e^2 {\rm
e}^{k|z|}}-\frac{k^2}{4(1+e^2 {\rm
e}^{k|z|})^2}-\frac{k}{1+e^2}\delta(z) \Big) \psi =0.
\label{nonextreme} \ee
Note that $m^2$ is the eigenvalue of the four-dimensional
Laplacian acting on $M(t,x_i)$. For the massless mode, this
reduces to a Schr\"{o}dinger-type wave equation with the solution
given by
\be \psi=N(e^2+{\rm e}^{-k|z|})^{1/2}. \label{solution} \ee
For the extremal case, this solution is square-normalizable and
the massless graviton is therefore bound to the braneworld
\cite{cvetic}. Away from extremality, the solution diverges for
large $z$ and is no longer physical. However, there is a
quasi-localized massive graviton mode, whose mass increases away
from extremality. This is illustrated numerically in Figure 2, by
a massive resonance on the braneworld. Notice that the wave
function dissipates further away from the extremal case, which
implies a decrease in the distance that the massive graviton
propagates on the braneworld before escaping into the extra
dimension.

\section{Braneworld from global $AdS_5$}

The RS2 braneworld model used a warped five-dimensional metric
that could be identified as a slice of $AdS_5$. Albeit the imposed
$\mbox{\elevenmsb Z}_2$ symmetry, this can be obtained from the
near-horizon region of the D3-brane reduced over $S^5$.

Consider the metric for $AdS_5$ expressed in global coordinates,
which is given by
\be ds_5^2=-{\rm cosh}^2 \rho\, dt^2+d\rho^2+{\rm sinh}^2 \rho\,
dx_i^2. \label{global} \ee
Imposing $\mbox{\elevenmsb Z}_2$ symmetry about $z$ we can express
(\ref{global}) as
\be ds_5^2=\frac{1}{{\rm sinh}^2(1+k|z|)}(-{\rm
cosh}^2(1+k|z|)dt^2+dx_i^2+dz^2), \label{global2} \ee
where ${\rm sinh}\, \rho=1/{\rm sinh}(1+k|z|)$.

We shall show that there is a massless graviton mode localized on the
braneworld. We insert the graviton
wavefunction $\Phi=\phi(z) M(t,x_i)$ into (\ref{min}) for the
background given by (\ref{global2}) and find the radial wave
equation to be
\be -\frac{{\rm sinh}^3(1+k|z|)}{{\rm cosh}(1+k|z|)}\partial_z
\frac{{\rm cosh}(1+k|z|)}{{\rm sinh}^3(1+k|z|)}\partial_z \phi=m^2
\phi, \label{wave} \ee
where $m$ is defined as in the previous sections. With the wave
function transformation
\be \phi=\Big( \frac{{\rm cosh}(1+k|z|)}{{\rm sinh}^3(1+k|z|)}
\Big)^{-1/2} \psi, \ee
the wave equation (\ref{wave}) can be expressed in Schr\"{o}dinger
form,
\be -\partial_z^2 \psi +V(z)\psi=m^2 \psi, \ee
with
\be 
V(z)=\frac{4\, {\rm sinh}^4 (1+k|z|)+20\, {\rm sinh}^2
(1+k|z|)+15} {{\rm sinh}^2 (2(1+k|z|))}k^2-\alpha\, k\, \delta
(z), \label{waveeqn}
\ee
where $\alpha=2(2{\rm sinh}^2(1)+3)/{\rm sinh}(2)$. This is a
volcano-type potential. That is, the coefficient of the
delta-function term is negative, and the $k^2$ term is maximum at
$z=0$ and decreases to $V(z\rightarrow \infty)=k^2$. The latter property
indicates that there is a mass gap of $M_{\rm gap}=k$ separating the 
localized massless state from the massive modes which propagate in the
extra dimension \footnote{Note that (\ref{waveeqn}) is the same graviton
wave equation as for D3-branes distributed uniformly over a disc
\cite{sfetsos}.}. 

The massless wavefunction solution is given by
\be \psi=N\sqrt{\frac{{\rm cosh}(1+k|z|)}{{\rm sinh}^3 (1+k|z|)}},
\label{massless} \ee
where $N$ is the normalization constant. Since this wavefunction
is square normalizable, it corresponds to a localized massless
graviton state. 

In order to see if the original RS2 scenario can arise within a
certain limit of the present braneworld scenario, we re-express
the metric for global $AdS_5$ (\ref{global}) via the coordinate
transformation ${\rm sinh}\, \rho=1/{\rm sinh}(\gamma +k|z|)$.
Rather than setting $\gamma =1$, we consider the case where
$\gamma$ is a small constant. In the limit $\gamma +k|z| \ll 1$,
the geometry asymptotes to a local slice of $AdS_5$. Therefore, in
the vicinity of our braneworld, we expect the physics to be
similar to that of the original RS2 scenario. Indeed, in this
limit with $\gamma$ replacing the ``1" in the massless graviton
wavefunction, (\ref{massless}) reduces to $\psi=N(\alpha
+k|z|)^{-3/2}$, which is the massless wavefunction for the RS2
scenario for this particular choice of coordinates.

Further away from the braneworld, the present scenario differs
quite drastically from the RS2 scenario. This is demonstrated by
the speed of gravitons in the bulk, which is $v(z)= {\rm
cosh}(1+k|z|)/{\rm cosh}(1)$, after a rescaling of $t$. Since
$v(z)$ increases away from the braneworld, there is an apparent
violation of causality. That is, as discussed in the Introduction,
gravitational disturbances may bend into the bulk and arrive at a
particular location on the braneworld earlier than does the light
from the same source.

\section{Discussion}

We have found that the breaking of global Lorentz invariance for a
braneworld may stem from various M-theoretic causes. One
possibility is that the spatial dimensions of a braneworld
correspond to relative transverse dimensions in the near-horizon
region of a system of intersecting $p$-branes. Another higher-dimensional
origin is an M5-brane in a constant $C$-field. For such
braneworld scenarios, the massless graviton, and any other massless
particle with corresponding bulk modes, is {\sl quasi}-localized with an
escape width which increases with momentum. Such metastability may
have phenomenological consequences, for example, for ultra-high
energy cosmic rays. As we discussed, another consequence in such a
scenario is a modification of the dispersion relation.

We next discussed the braneworld scenario which results from the
near-horizon region of a $p$-brane that is perturbed away from
extremality. In particular, we considered the semi-analytically
tractable example of a 5-brane in heterotic or type II theories.
The global breaking of Lorentz invariance in the corresponding
braneworld gives rise to a {\sl quasi}-localized graviton which is
massive. As the corresponding 5-brane is perturbed further from
extremality, the mass of the quasi-bound state increases, and the
graviton propagates for a decreasing distance on the braneworld
before escaping into the bulk. Another odd consequence, which was
discussed in \cite{jvp2}, is the presence of quasi-bound massive
harmonic states, which travel a relatively short distance on the
brane before escaping.

We considered a braneworld scenario which arises from global
$AdS_5$. Unlike the case of a local patch of $AdS_5$, which was
used in the original RS2 scenario, Lorentz invariance is globally
broken on the brane since the speed of gravitons increases away
from the braneworld. As discussed in the Introduction, the result
of this is that gravitons can travel from one point of the
braneworld to another by bending into the bulk, and thus arriving
apparently faster than light.

It seems that, in order to take the Penrose limit of $AdS_5 \times
S^5$, $AdS$ must be expressed in global coordinates.\footnote{This
results in a background on which the type IIB light-cone string
action is exactly solvable \cite{blau1, blau2, metsaev,
berenstein}.} Therefore, there may be an interesting connection
with the braneworld scenario arising from $AdS_5 \times S^5$, for
which $AdS$ is in global coordinates. In taking this limit, one
considers particles traveling fast around a coordinate in $S^5$.
Therefore, one would need to consider gravitational modes which
move fast around $S^5$ \cite{rotational}.

Lastly, since this falls under the category of gravitational Lorentz
violations from M-theory, we would like to make mention that
five-dimensional $AdS$-Reissner-Nordstrom black holes have the
ten-dimensional interpretation of spinning D3-branes \cite{duff}. This
enables us to place previous work on gravitational Lorentz violations
within the context of ten dimensions and holography, in particular with
regards to the Coulomb branch of the dual gauge theory. Future work will
elaborate on this point.

\section*{Acknowledgments}

We are grateful to Malcolm Fairbairn for helpful conversations on
cosmological issues and to Don Marolf for a useful discussion on
lightcone structures. J.F.V.P. thanks Hong L\"{u} and Jack
Gegenberg for useful correspondence and discussions.

\end{document}